# Pivoting as an Adaptive Strategy to Geopolitical Tensions in U.S. Science

Moxin Li [1], Yifang Ma [2], Yang Wang [1,*], and Dashun Wang [3,4,5,6,7,*]

1 School of Public Policy and Administration, Xi'an Jiaotong University, Xi'an 710049, China

2 Department of Statistics and Data Science, Southern University of Science and Technology, Shenzhen 518055, China

3 Center for Science of Science and Innovation, Northwestern University, Evanston, IL 60208

4 Kellogg School of Management, Northwestern University, Evanston, IL 60208

5 Ryan Institute on Complexity, Northwestern University, Evanston, IL 60208

6 Northwestern Innovation Institute, Northwestern University, Evanston, IL 60208

7 McCormick School of Engineering, Northwestern University, Evanston, IL 60208

* To whom correspondence may be addressed. Email: yang.wang@xjtu.edu.cn, or dashun.wang@northwestern.edu.

**Abstract**

Geopolitical tensions increasingly reshape the structure and openness of global science, yet we still lack a clear understanding of how successfully scientists adapt their work under such pressures. Using millions of funding and publication datasets across the past ten years, we investigate how U.S.-China geopolitical tensions reshaped individual research activities of U.S.-based scientists, particularly those collaborating with Chinese peers. We find that although U.S.-China geopolitical tensions significantly reduce funding opportunities, many scientists actively respond by pivoting their research portfolios toward alternative topics, and this adaptive reorientation partially mitigates funding losses. Crucially, the effectiveness of this adaptive strategy is highly unequal: for scientists in high-risk domains, those of Asian descent, and early-career scientists, pivoting offers only limited protection against funding loss. Our results demonstrate that geopolitical tensions reshape science through shifts in scientists' strategic decisions about their research focus. Understanding this adaptive but uneven reconfiguration is essential for science policies to strengthen the resilience and inclusiveness of the scientific enterprise.

**Introduction**

Complex interplays among governments, scientific communities, and broader society have increasingly amplified the influence of politics on scientific endeavors, particularly through shifts in funding priorities and policy directives (1-6). Among recent national and international initiatives (7-10), the U.S. Department of Justice's "China Initiative" - a policy brought in by the Department of Justice to counter perceived economic espionage and foreign influence – stands out as a vivid example of how political events disrupt the structure and openness of science, including international collaborations, knowledge production, psychological toll, data sharing initiatives (10-20). Although existing studies document how such policies disrupt scientific openness and generate uncertainty for affected researchers, much less is known about whether and how scientists actively adjust their research in response. Understanding these adaptive behaviors has profound implications for both individual careers and the broader trajectories of U.S. science (21-24).

Prior work examining geopolitical influence on science has followed two distinct regimes. One line investigated institutional collapse and forced displacement, where scientific institutions fragment and talent is physically redistributed (25, 26). A second line focused on recent U.S.-China tensions, showing both reduced productivity and heightened fear, uncertainty, and withdrawal among researchers (14, 18). In both regimes, scientists are principally portrayed as passive recipients of geopolitical shocks, either uprooted by institutional breakdown or constrained by external scrutiny. Here, we study a fundamentally different regime: adaptive response under constraint. In this setting, the institutional structure of U.S. science remains formally intact, yet geopolitical pressures quietly but powerfully reshape the incentives and perceived risks associated with specific research domains. This environment allows us to move beyond disruption to examine how scientists actively adapt their research portfolios in place, revealing a mechanism of adjustment that precedes, and may prevent, system-level collapse.

Although adaptive responses occur at the level of individual scientists, the scale of such activity hinges on fundamental trade-offs and uncertainties. Scientists must decide whether to make incremental or substantial shifts in their research directions to sustain long-term career trajectories. Anecdotal evidence - for instance, MIT professor Gang Chen's strategic shift from semiconductor research to fluid dynamics - suggests that scientists may reorient their research in response to geopolitical tensions (12). At the same time, prior studies highlighted that the "burden of knowledge", resource constraints, and current funding designs can substantially constrain scientists' capacity to rapidly redirect their research agendas (27-29). Crucially, venturing into new domains often entails a "pivot penalty", characterized by reduced citation impact and elevated professional risk (23, 30). These mechanisms create a central tension: geopolitical pressures heighten incentives to adapt, yet cognitive and structural frictions constrain scientists' ability to do so.

The emergence of large-scale scholarly datasets has enabled us to examine individual scientists' behaviors at unprecedented resolution (31-38). Using millions of funding and publication records over the past decade, we investigate how scientists actively reconfigure their work in response to geopolitical tensions. Specifically, we address two central questions: (1) How do U.S.-China geopolitical tensions affect funding and topic selection among U.S.-based scientists? And, (2) to what extent do pivoting behaviors serve as adaptive strategies to mitigate funding uncertainties? Using a recently developed framework to quantify pivoting behaviors (23), we find that scientists actively pivot to new topics in response to U.S.-China geopolitical tensions. We further find that pivoting behaviors play a strong moderating role, alleviating some funding losses associated with geopolitical tensions. Finally, our results indicate that the effectiveness of pivoting as an adaptive strategy is highly uneven across different groups, offering limited protection for scientists in high-risk research domains, those of Asian descent, and early-career researchers.

Together, these findings reveal how science adapts when openness meets geopolitics. By uncovering both the behavioral mechanisms of adaptation and the unequal distribution of adaptive capacity, our work advances a framework for understanding how geopolitical tensions reshape scientific careers, research agendas, and the evolution of knowledge within an intact scientific system.

**Research Design**
**Cohort Selection and Measurement Framework.** We use the Dimensions dataset, one of the largest and most comprehensive scholarly dataset, containing approximately 140 million publications, and 7.1 million grants (39, 40). Here, we focus on U.S.-based scientists working in science, technology, engineering, and mathematics (i.e., STEM fields, see Supplementary Information Section 1.1). Figure 1A illustrates the cohort selection process (see Data and Cohort Selection in Methods). We begin by identifying STEM scientists who engaged in international collaborations between 2009 and 2013. For the treatment group, we select U.S.-based scientists who co-authored at least three papers with Chinese scientists during this period. We then refine this group by retaining only those who published at least three papers between 2014 and 2018. Among them, we restrict the cohort to those who served as principal investigators (PIs) or co-principal investigators (co-PIs) on at least one grant, yielding a final treatment group of 13,104 U.S.-based scientists with prior collaborations with Chinese scientists. For the control group, we apply similar selection criteria to U.S.-based STEM scientists who collaborated exclusively with countries other than China, resulting in a final sample of 52,059 scientists. Note that similar cohort selection strategies have been used in recent studies (14, 15). To ensure comparability between the treatment and the control group prior to geopolitical tensions, we apply the Coarsened Exact Matching based on pre-treatment characteristics (i.e., CEM, see Matching Strategy in Methods). Our matched sample comprises 6,926 scientists in the treatment group and 13,685 scientists in the control group.

We calculate each scientist's annual funding amount by aggregating grants awarded to them as a PI or co-PI (see Funding and Pivot Size in Methods). Figure 1B shows an illustration example of one scientist, and s/he obtained three grants from 2015 to 2021 as a co-PI (Figure 1B). To quantify pivoting behaviors at the individual level, we compare reference journal composition of a given work and a scientist's prior work using a cosine-similarity metric (Figure 1C, and Funding and Pivot Size in Methods) (23). Table 1 Panel A reports the descriptive statistics for the matched sample. We observe no statistically significant differences between the treatment and the control group across any dimension prior to treatment, supporting the validity of the matching procedure. The distribution of key covariates for both groups is further illustrated in Supplementary Information Figure S3. This absence of pre-treatment differences suggests that any difference observed after 2019 can be attributed to the U.S.-China geopolitical tensions, rather than to pre-existing differences.

**Results**
**Attrition in response to geopolitical tensions.** We first examine scientists' survival rates following the onset of geopolitical tensions. We define attrition as the absence of any publishing activity for three consecutive years (see Supplementary Information Section 1.2). We compute survival rates for scientists in both the treatment and control groups from 2014 to 2023, and use the Kaplan–Meier estimator to compare differences in survival probabilities between these groups over time.

Figure 2A shows that prior to 2019, there is no significant difference in terms of survival rates between the treatment and the control group (Log-rank test, $p$-value = 0.894). However, after 2019, the treatment group exhibits a pronounced decline in survival rates. By the end of 2021, the survival rate drops to 92% in the treatment group, while the control group maintains a rate of 95% (Log-rank test, $p$-value < 0.001). Notably, 64% of those who stopped publishing within the treatment group are of Asian descent, compared to 26% in the control group. This disparity is especially striking given that Asian scientists constitute 56% of the treatment group overall ($X^2$ test, *p*-value < 0.001). In contrast, the fraction of Asian scientists in the control group mirrors the proportion among those who left academia ($X^2$ test, *p*-value = 0.086). These findings suggest that Asian scientists are particularly vulnerable to U.S.-China geopolitical tensions (14, 18). Moreover, we find that attrition is disproportionately concentrated in scientific domains closely aligned with U.S.-China strategic competitions, such as Engineering & Computing Science and Mathematical & Physical Sciences (18) (Figure 2C).

**Funding and pivot size.** We next assess the funding acquisition and pivoting behaviors among U.S.-based scientists after the U.S.-China geopolitical tensions. We first show descriptive results in Table 1 Panel B, and we find that scientists in the treatment group experienced significantly larger declines in total funding amount (*p*-values = 0.005), U.S. funding amount (*p*-values < 0.001), U.S. federal funding amount (*p*-values < 0.001) and exhibited a stronger tendency to switch research topics (*p*-values = 0.001).

Building on these descriptive results, we show in Table 2 the baseline regression results. Columns 1 and 2 show that the difference-in-differences (i.e., *DID*) coefficients for U.S. federal funding are significantly negative, indicating that scientists in the treatment group experienced significant reductions in obtaining U.S. federal funding compared to their counterparts. After controlling for confounders, scientists in the treatment group received, on average, 17.5% less annual U.S. federal funding compared to those in the control group (($e^{-0.193} - 1$)*100%, *p*-value = 0.001). Columns 3 and 4 further show that scientists in the treatment group showed a 17.1% decline in U.S. funding (*p*-value = 0.002) and a 10.8% decline in overall funding (*p*-value = 0.059) compared to their counterparts. These patterns suggest that funding reductions were particularly concentrated among U.S. federal sources (41, 42).

Critically, we find that scientists actively pivoted from their usual research topics in response to the geopolitical tensions. Table 1 Column 6 shows that treatment group scientists were significantly more likely to pivot away from their usual research topics compared to the control group (*p*-values < 0.001). We obtain consistent results using the event study analysis (Figure 3). Specifically, Figures 3A-C show that scientists in the treatment group showed significantly less U.S. federal funding, U.S. funding, and total funding following the U.S.-China geopolitical tensions, respectively. Importantly, we also observe a significant increasing trend in the treatment group's tendency to switch research topics. This upward trend emerged in 2019, suggesting that many scientists proactively adjusted their research portfolios in response to geopolitical tensions (Figure 3D).

The negative effect of the geopolitical tensions remains robust across alternative settings. To assess scientists' ability to secure funding, we use the number of grants awarded to each scientist; to quantify the extent of topic shifts, we use the fraction of overlapping references between a scientist's current and prior publications (see Supplementary Information Section 4.2). We find that the geopolitical tensions significantly reduced the total number of U.S. federal grants (Column 7, *p*-value = 0.005), the number of U.S. grants and overall grants (Columns 8 and 9, *p*-value = 0.009 and *p*-value = 0.039, respectively). Additionally, treatment group scientists published papers with significantly lower reference overlap relative to their prior work compared to their contenders (Column 10, *p*-value < 0.001), indicating substantial shifts in their research focus.

We conduct additional analyses to address potential confounding effects between funding and pivoting behaviors (Supplementary Information Section 5). Figure S4 shows that scientists in the treatment group are more likely to switch topics before receiving any new grant after geopolitical tensions, indicating that pivoting is not solely driven by funding (Supplementary Information Section 5.1). To further validate this finding, we replicate the regression analysis using only papers published after the geopolitical tensions but before the first new grant after geopolitical tensions, finding consistent results (Table S2). Finally, consistent with prior work (14, 23), we also find that treatment group scientists experienced a sustained decline in citation impact and that pivoting is associated with lower citation impact (see Supplementary Information, section 5.2 and 5.3). Taken together, our findings reveal a robust empirical pattern: scientists reduced access to funding after geopolitical tensions, but they actively adjusted their research portfolio in response to geopolitical tensions.

**The Moderating Role of Pivoting Behaviors**. Our findings reveal that U.S. scientists who had prior collaborations with Chinese scientists experienced significant challenges in securing funding and exhibited a marked tendency to shift their research topics. Such topic shifts, however, come with inherent costs: the "burden of knowledge" makes it difficult for scientists to generate valuable insights outside their established domains (29), while pivot penalty can hinder the recognition of their work by new audiences (23, 43). Despite these challenges, U.S. scientists collectively adjusted their research portfolios in response to geopolitical tensions. To further inform the adaptive role of the pivoting behaviors, we next investigate whether pivoting behaviors moderate funding declines caused by geopolitical tensions.

We introduce an interaction term between pivot size and the original *DID* estimator. Table 3 demonstrates that pivoting behaviors significantly moderate the decline of funding after the geopolitical tension. In Column 1, after controlling for all covariates, scientists who switch topics experienced a significantly smaller decline in U.S. federal funding ($p$-value = 0.002). Columns 2 and 3 show that pivoting behaviors also mitigate declines in both U.S. funding and total funding ($p$-value = 0.003 and 0.007, respectively). Columns 4 to 6 examine the number of grants received, and the interaction terms remain positive and statistically significant, particularly for U.S. federal grants ($p$-value < 0.001). To further validate this finding using alternative measurements of pivot size, we introduce an interaction term between the *DID* estimator and the reference overlap in Supplementary Information Table S5, finding consistent results (all $p$-values < 0.001), indicating that scientists who remain closer to their prior research topics experienced greater funding losses after the geopolitical tensions.

To provide additional evidence, we divide scientists into two groups according to their pivot sizes. We find that treatment-group scientists who continued to pursue their prior topics experienced more funding decline after the geopolitical tensions compared to their counterparts in the control group (Supplementary Information, Figure S5). In contrast, among those who switched research topics, we observe no significant difference in terms of U.S. federal funding and U.S. funding between the treatment and control groups (Supplementary Information, Figure S5A-B, $p$-value=0.585 and 0.548, respectively). Notably, in terms of total funding amount, treatment-group scientists who pivoted to new topics showed slight advantages (Supplementary Information, Figure S5C, $p$-value=0.007). Taken together, these results provide compelling evidence that pivoting behaviors can help mitigate funding declines induced by the geopolitical tension, highlighting the critical role of this adaptive strategy.

**Effects of field, race, and career stages.** We examine the effect of scientific domains, racial identities, and career stages. First, we investigate field-specific risks by categorizing scientific fields into high- and low-risk domains based on the median value of the attrition rate (Supplementary Information section 5.5, Figure S6). High-risk fields – including particle and high energy physics, nanotechnology, electronics sensors and digital hardware – closely align with domains identified as geopolitically sensitive in policy reports (44, 45). We find that scientists in these high-risk domains experienced more pronounced declines in funding and were more likely to shift research topics in response to geopolitical tensions (Supplementary Information, Figure S7, Table S6 columns 1-8). Crucially, Figure 4A-B show that pivoting behaviors fail to mitigate funding declines associated with the U.S.-China geopolitical tensions in high-risk domains and are associated with even larger funding losses in these fields ($p$-value = 0.016 for U.S. federal funding and $p$-value = 0.017 for U.S. funding). In contrast, only in low-risk fields do pivoting behaviors significantly mitigate funding declines ($p$-value < 0.001), suggesting that geopolitical tensions may constrain the adaptive capacity of scientists in certain disciplines. Note that the regression results are reported in Supplementary Information Table S6.

Next, we explore the distinct challenges faced by Asian scientists. We identify authors' ethnicity using the *rethnicity* package, which uses a character-level bidirectional long short-term memory (BiLSTM) model (46). We find that both Asian and non-Asian scientists experienced significant reductions in terms of funding amount followed by the geopolitical tensions (Supplementary Information, Table S7 columns 1-6). We also find that there is no significant difference in pivoting behaviors between the treatment and the control group among Asian scientists (Supplementary Information, Figure S7K; Table S7 columns 7, and 8). This pattern may reflect the influence of stereotype threat, whereby Asian scientists, even without direct collaborations with Chinese counterparts, feel heightened pressure to adjust their research portfolios in anticipation of increased scrutiny related to the U.S.-China geopolitical tensions (18). Yet, as shown in Figures 4C and D, pivoting behaviors fail to confer protective effects on funding for Asian scientists ($p$-values = 0.192 and 0.183 for both U.S. federal and U.S. funding). This disparity highlights potential barriers that limit the effectiveness of adaptive strategies among Asian scientists.

Finally, we examine the effect of career stages by dividing scientists into junior and senior cohorts based on the median year of their first publication. Senior scientists experienced larger funding losses in response to geopolitical tensions (Supplementary Information, Table S8 columns 1-6). This pattern is consistent with prior survey evidence that senior scientists were less likely to apply for federal funding after the geopolitical tensions (18). We also find suggestive evidence that junior scientists display a greater tendency to switch topics (Table S8 columns 7, and 8). However, we find that only senior scientists benefit from pivoting behaviors in terms of mitigating funding losses associated with the U.S.-China geopolitical tensions (Figure 4E-F; $p$-value = 0.005 for U.S. federal funding and $p$-value = 0.005 for U.S. funding). Taken together, our analysis reveals that the adaptive role of pivoting behaviors substantially vary for different scientists. Scientists in high-risk fields, those of Asian descent, and junior scientists appear disproportionately affected by the U.S.-China geopolitical tensions. These findings highlight the need for targeted policy interventions to support equity and resilience within the scientific community.

**Discussion**

Modern science has expanded rapidly since the seventeenth century under a core principle of openness, advancing as a collective enterprise that continually pushes the boundaries of knowledge. Over the past several decades, this openness has been exemplified by a surge in international collaborations, including landmark achievements such as the Human Genome Project, the discovery of gravitational waves, and the Paris Agreement on climate change (47-49). This foundational principle is increasingly strained under recent geopolitical tensions. The findings in this study, however, highlight that scientists are not passive recipients of these geopolitical tensions. Instead, many respond by reconfiguring their research portfolios toward alternative topics, and this adaptive behavior partially mitigates funding losses associated with geopolitical tensions, suggesting that individual behavioral flexibility serves as a mechanism of resilience within an otherwise stable scientific system. Our findings highlight that when openness meets geopolitics, science does not simply contract or collapse; it actively reorganizes, though not without consequence.

We observe that the benefits and costs of the adaptive behaviors are unevenly distributed across fields, career stages, and demographic groups, suggesting emerging inequalities in who can adapt, how, and at what price. In particular, we observe that Asian scientists are more likely to shift their research focus in response to geopolitical tensions - even in the absence of direct collaboration with Chinese scientists - suggesting the presence of a stereotype threat and risk sensitivity. Junior scientists and scientists in high-risk domains also face disproportionate trade-offs, given their more limited buffers of funding access. These patterns highlight the need for targeted institutional support for scientists who are disproportionately exposed to geopolitical tensions yet are essential to American science - support that is crucial for sustaining a resilient and inclusive scientific ecosystem.

Pivoting also carries inherent trade-offs, which have critical implications for individual scientists who must weigh the relative risks of remaining in usual research areas against the uncertainties of entering new ones. While our results show that pivoting may help preserve funding opportunities, prior research has cautioned that shifting research directions can incur a "pivot penalty" - manifesting in reduced citations, reduced intellectual continuity, or challenges in building credibility in new fields (23). This dilemma is particularly salient in the current hyper-competitive funding environment, where access to grants is a critical determinant of both scientific progress and career advancement (28, 50, 51).

By identifying pivoting as a form of adaptive reorientation rather than simple contraction or withdrawal, our results introduce an agency-centered perspective into the geopolitical studies of science. Existing work has emphasized how external shocks reshape collaboration structures or knowledge production (14-16), yet has paid limited attention to how individual scientists respond when facing these altered conditions. Our results show that scientists can restructure their research trajectories to maintain funding access, demonstrating that scientific systems possess a degree of endogenous resilience rooted in scientists' decision-making rather than institutional reform alone.

The observed adverse effects on funding likely operate through multi-dimensional underlying mechanisms. Scientists with prior collaborations in China may proactively avoid applying for new U.S. federal funding, as suggested by recent survey evidence (14, 18). Moreover, even when such scientists do apply, potential biases in the peer review or evaluation process caused by geopolitical tensions may further restrict their access to funding (52). These results also carry important implications for U.S. funding agencies who must critically examine and address potential implicit biases, ensuring that grant decisions are based on scientific merit rather than the national origins.

Our findings also resonate with growing concerns about the instability of scientific funding in the United States, as the U.S. enters a new era marked by fiscal constraints and shifting political priorities (6, 53-55). Geopolitical tensions, particularly with China, appear to exacerbate existing structural vulnerabilities in the U.S. scientific ecosystem. When funding uncertainty intersects with geopolitical tensions, the challenges facing scientists become especially pronounced, amplifying existing pressures on U.S. science. This confluence of risks may contribute to an inverse brain drain, as an increasing number of U.S.-based scientists seek opportunities outside the U.S. (56, 57). Such a trend poses a serious threat to the long-term capacity of the United States to sustain its leadership in scientific discovery and innovation.

Finally, it is worth noting that although the Dimensions dataset offers among the most comprehensive coverage of grant records, it has limitations in identifying grants originating from academic institutions and industry, and its coverage outside the U.S. may be incomplete. Nevertheless, our analysis primarily focuses on U.S. federal funding, which mitigates these concerns. Importantly, the implications of geopolitical tensions may extend beyond the U.S. scientific community, reshaping global science in lasting ways, including scientific mobility, international collaborations, and innovation across public and private sectors. As former U.S. Secretary of Energy Steven Chu said, "You won't see the impact on U.S. science and technology for at least a decade". Indeed, the long-term consequences of geopolitical tensions may unfold gradually, influencing both the structure and trajectory of knowledge production for years to come. Future research should explore the broader and longer-term impacts of geopolitical tensions across a wider range of outcomes. As the global scientific workforce becomes increasingly intertwined with geopolitical forces, understanding these relationships is critical for designing resilient and inclusive science policies.

**Materials and Methods**
**Data and Cohort Selection.** We use the Dimensions dataset to extract publication metadata, including titles, publication venues, author names, affiliations, publication dates, scientific domains, references, and funding sources. Dimensions uses the Australian and New Zealand Standard Research Classification system, which organizes science into 22 broad fields and 154 subfields through a two-tier Field of Research (FOR) framework, facilitating granular analysis of disciplinary landscapes (58). Dimensions provides one of the most extensive global records of scientific funding, which cover over 7.1 million grants/$2.36 trillion awarded by over 700 funding agencies worldwide (Figure S1) (59). The dataset's quality and scope have been widely recognized in prior research, and it has been used extensively for systematic analyses of global funding landscapes (7, 23).

We focus on U.S.-based STEM scientists for the following reasons (see Supplementary Information, Section 1.1). First, STEM research is central to the scientific agendas of both China and the United States, which ensures a balanced disciplinary composition between the treatment and the control group (60, 61). Second, STEM research often relies heavily on international collaborations, and the United States and China have developed particularly strong partnerships in these domains over the past two decades (15, 44). Third, STEM fields are at the center of U.S.-China geopolitical tensions, as China's growing capabilities raise U.S. concerns over competitiveness and national security. These concerns lead to visa restrictions, funding scrutiny, and barriers to collaboration, making STEM a primary site where geopolitical tensions shape basic research (62).

**Funding and Pivot Size.** We calculate funding amount and pivot size for each scientist. For each grant, we divide the total award equally among all listed PIs and co-PIs, and then distribute each investigator's share evenly across the grant's active years. If a scientist holds multiple grants in a given year, we sum all corresponding funding shares to obtain his/her total annual funding (Figure 1B). To normalize the distribution, we apply a logarithmic transformation to each scientist's annual funding amount. Based on the funding source, we classify each scientist's funding into U.S. funding and U.S. federal funding (see Supplementary Section 3). We show the distributions of funding amount in Supplementary Figure S2 A–C.

When calculating pivot size, for each paper $p$ published at time $t$, we calculate the cosine distance between its reference journal composition vector $V_i^p$ and a baseline vector $V_i^{baseline(t)}$, which aggregates all papers published in an earlier period (Figure 1C):

$$\phi_i^t = \frac{1}{N}\sum_p \left(1 - cos(V_i^p, V_i^{baseline(t)})\right).$$

Specifically, for papers published between 2014 and 2018, we construct the baseline vector using papers published between 2009 and 2013; likewise, for papers published between 2019 and 2023, we calculate the baseline vector using papers published between 2014 and 2018:

$$V_i^{baseline(t)} = \begin{cases} \sum_p V_i^{p \in [2009, 2013]}, & \text{if time } t \text{ is between 2014 and 2018} \\ \sum_p V_i^{p \in [2014, 2018]}, & \text{if time } t \text{ is between 2019 and 2023} \end{cases}$$

Where $p$ represents papers published by scientist $i$ in year $t$, and $N$ is the total number of papers by scientist $i$ in that year. We then average the pivot size for each scientist across all their papers each year to obtain their annual pivot size $\phi_i^t$. Supplementary Information Figure S2D presents the distribution of the topic pivot size for both the treatment and the control group.

**Matching Strategy.** To ensure comparability between the treatment and the control group prior to geopolitical tensions, we apply the Coarsened Exact Matching method. Scientists are matched based on a set of pre-treatment characteristics, including scientific domains, career stage, U.S. federal funding amount, total U.S. funding amount, total funding amount, pivot size, annual publication count, average normalized citations, and the number of new U.S.-based collaborators (see Supplementary Information, Section 4.1). These covariates are included to ensure that treated and control scientists are comparable in their career stage, scientific domain, their baseline capacity to secure funding, their tendency to switch research topics, their productivity, citation impact, and their U.S.-based collaborative networks.This procedure results in a final matched sample comprising 6,926 scientists in the treatment group and 13,685 scientists in the control group. We show the distribution of several key covariates prior to treatment in Figure S3, and we find there is no significant difference between the treatment group and the control group.

**Regression Framework.** To account for potential confounding factors, we control for a set of covariates at both the individual and paper levels in the regression framework. These include career age (i.e., the number of years since a scientist's first publication), annual scientific productivity (i.e., the total number of papers the scientist publishes in a given year), and average normalized citation impact adjusted by field and year (see Supplementary Information Section 4.1) (63), and number of new U.S. collaborators (i.e., U.S.-based coauthors who had not previously collaborated with the focal scientist). Finally, we include individual and year fixed effects to account for time-invariant characteristics at both the scientist and year levels.

$$Y_{i,t} = \beta_0 + \beta_1 Treat_i \times Post_t + \gamma X_{i,t} + \delta Individual_i + \theta Year_t + \varepsilon_{i,t} \quad (1)$$

We use the standard difference-in-differences framework to estimate the impact of U.S.-China geopolitical tensions, as specified in equation (1). The variable *Treat* is a binary indicator equal to 1 for scientists in the treatment group and 0 for those in the control group. Geopolitical tension is set to begin in 2018, corresponding to the launch of the "China Initiative" by the U.S. Department of Justice in November of that year. Accordingly, the post-policy period is defined by the variable

*Post*, which is coded as 1 for years from 2019 onward and 0 otherwise (14). The *DID* estimator is captured by the interaction term *Treat×Post*, which isolates the differential effect of geopolitical tensions on the outcome of interest for the treatment group relative to the control group. In Equation (1) $Y_{i,t}$ represents either the annual funding amount or the pivot size for scientist $i$ at year $t$. $X_{i,t}$ represents control variables at year $t$, including career stages, scientific field, prior funding amount, prior pivot size, productivity, impact, and the number of new U.S. collaborators. $Individual_i$, and $Year_t$ represent individual fixed effects and year fixed effects, respectively. Standard errors are clustered at the individual level to correct for within-scientist correlations over time.

**Acknowledgments.** This work is supported by the National Natural Science Foundation of China (Grant No. 72004177). We thank Alexander Furnas, Lingfei Wu, Yi Bu, Chunliang Fan, and participants of the 2025 International Conference on Science of Science and Innovation (ICSSI) for their insightful comments and suggestions.

**Author Contributions:** M.L. did the primary statistical analysis. Y.W. and D.W. conceptualized and designed the research and reviewed all the results, Y.W. Y.M. and D.W. led in drafting the manuscript, and all authors contributed to editing the manuscript.

**Competing Interest Statement:** The authors declare that they have no competing interests.

# Figures and Tables

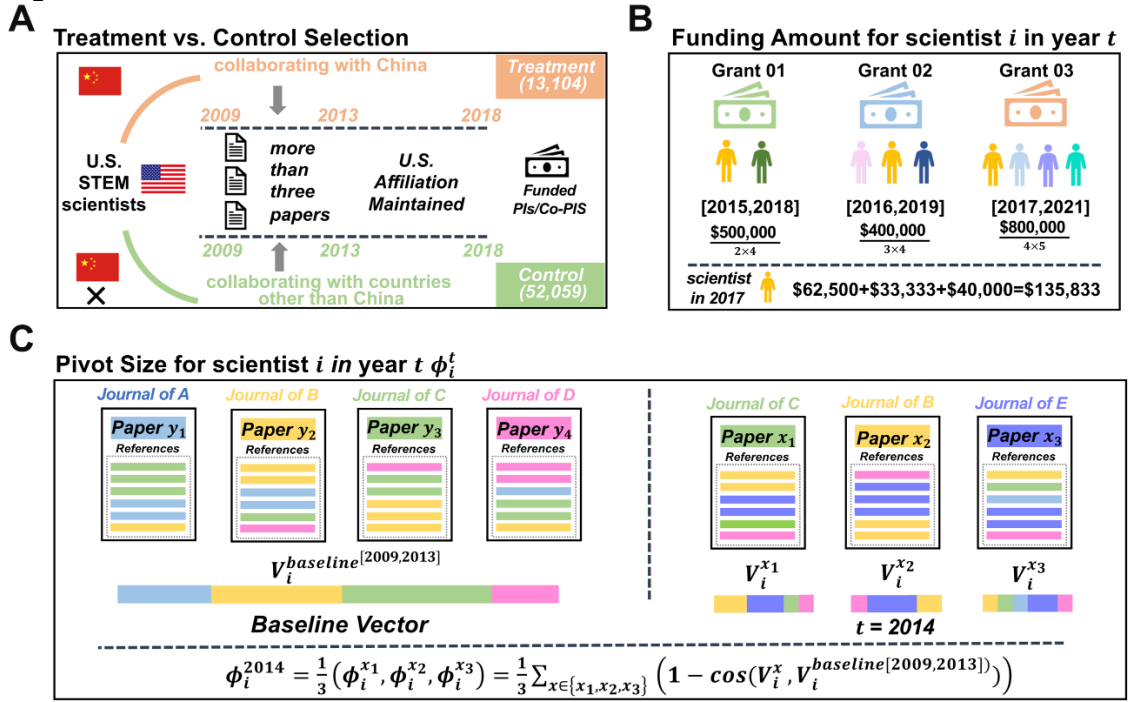

**Figure 1.** The cohort selection method and the process of calculating the annual funding amount and pivot size. (A) We define the treatment group as U.S.-based STEM scientists who collaborated with Chinese scientists between 2009 and 2013. The control group includes comparable U.S. STEM scientists whose international collaborations were with non-Chinese scientists during the same period. (B) Illustrates the calculation of annual research funding. For each grant where a scientist is listed as principal investigator (PI) or co-principal investigator (co-PI), we divide the total funding amount equally among all listed investigators and then distribute each investigator's share evenly across the grant's duration. Then, we sum all active grants per year. (C) shows how we measure pivot size. For each paper, we calculate the cosine similarity between the cited journals in its reference list and the cited journals in the reference list from the scientist's prior work. We then average the pivot size for each scientist across all their papers per year.

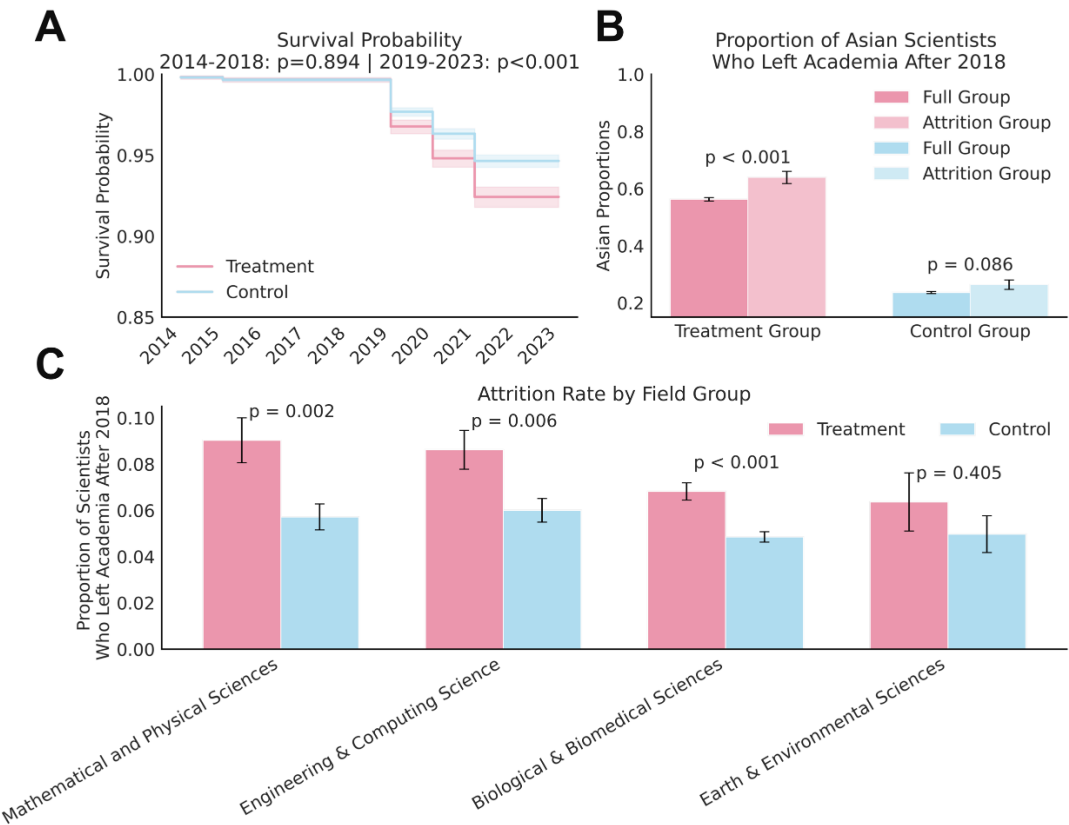

**Figure 2.** Attrition patterns between treatment and control groups. (A) Survival analysis (2014–2023). Attrition is defined as stopping publishing for at least three consecutive years. (B) Share of Asian scientists among those who exited academia after 2019 in the treatment and control group. (C) Attrition rate across various scientific domains after 2019.

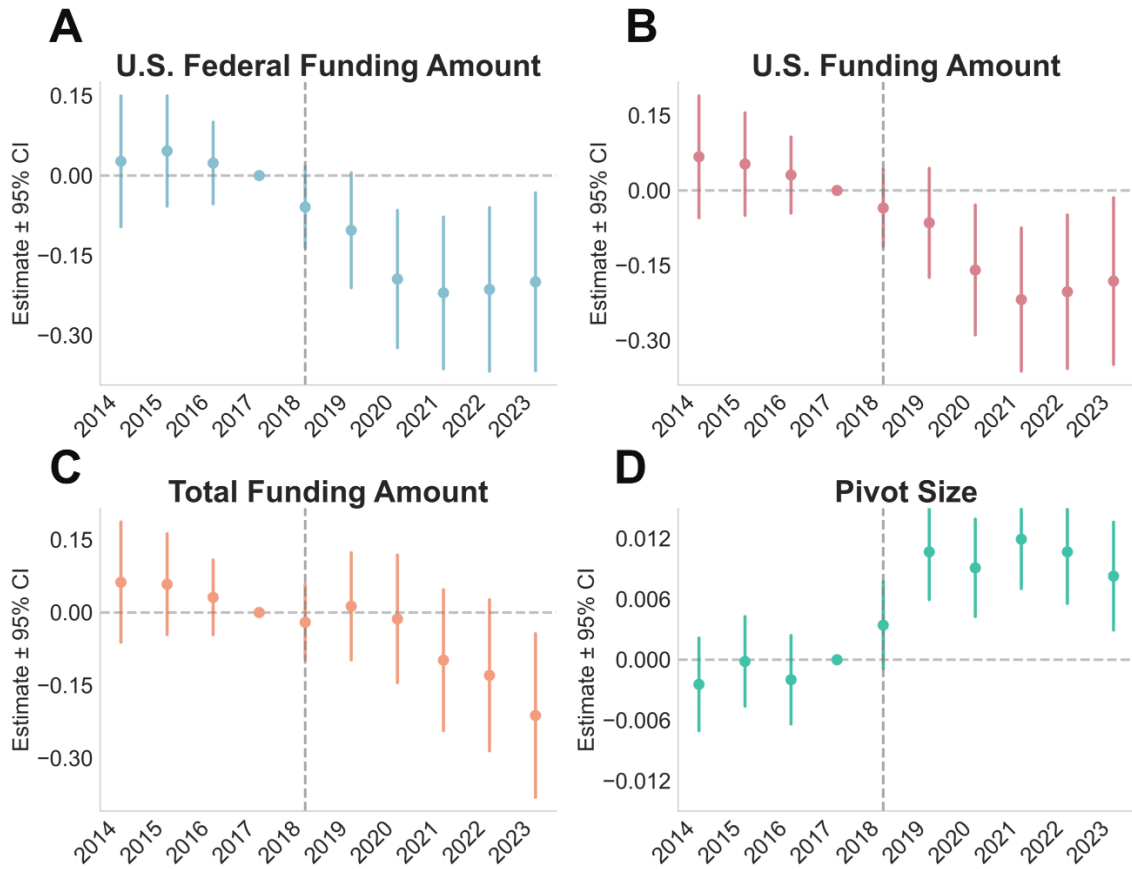

**Figure 3.** The event study analysis. (A) – (D) represent the U.S. federal funding amount, U.S. funding amount, total funding amount, and pivot size, respectively. Error bars represent 95% confidence intervals. In the regression analysis, we control for career age, number of publications, average normalized citations, and the number of new U.S. collaborators, author fixed effects, and year fixed effects in the regression analysis. Standard errors are clustered at the individual scientist level.

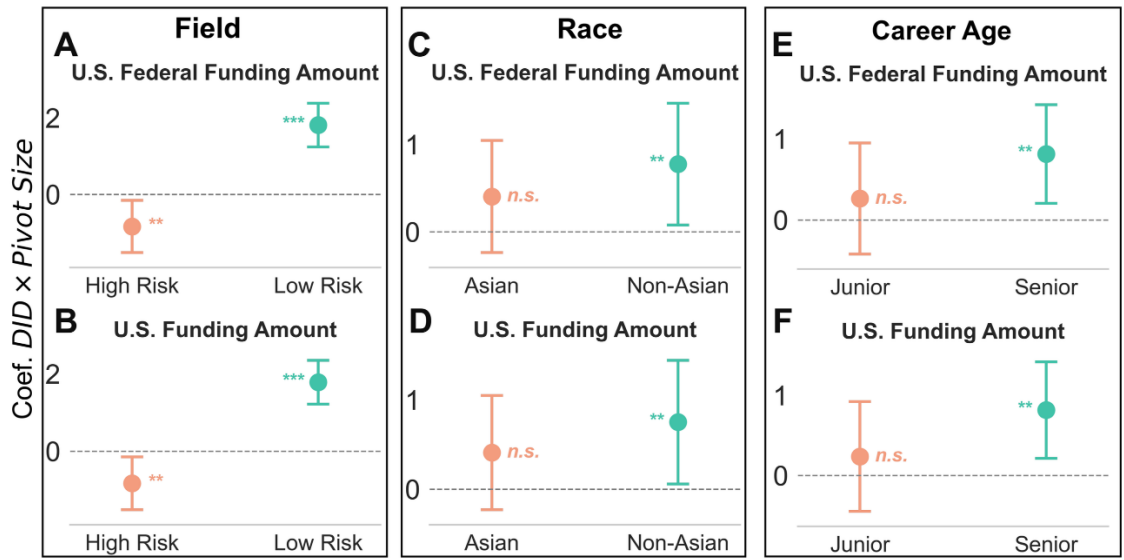

**Figure 4.** The effect of scientific domain, race, and career age on the moderating effect. (A) Estimated interaction effect on U.S. federal funding by field risk level. (B) Estimated interaction effect on total U.S. funding by field risk level. (C) Estimated interaction effect on U.S. Government funding by race. (D) Estimated interaction effect on total U.S. funding by race. (E) Estimated interaction effect on U.S. federal funding by career stage. (F) Estimated interaction effect on total U.S. funding by career stage. We show regression coefficients and their corresponding 95% confidence intervals. All regressions control for career age, number of publications, average normalized citations, and the number of new U.S. collaborators, author fixed effects and year fixed effects. Standard errors are clustered at the individual scientist level.

**Table 1.** Summary Statistics using the matched sample. Note that variables related to funding amount, number of publications, citation impact, and number of collaborators are log-transformed.

|  | Treatment Mean | Control Mean | p-value |
|---|---|---|---|
| **A. Pre-treatment (2014–2018)** | | | |
| U.S. Federal Funding Amount | 8.534 | 8.608 | 0.344 |
| U.S. Funding Amount | 8.678 | 8.755 | 0.322 |
| Total Funding Amount | 8.806 | 8.844 | 0.618 |
| Pivot Size | 0.507 | 0.505 | 0.479 |
| Career Age | 21.900 | 21.799 | 0.476 |
| # Papers | 1.907 | 1.912 | 0.639 |
| Average Normalized Citations | 0.643 | 0.639 | 0.329 |
| # New U.S. Collaborators | 2.234 | 2.255 | 0.159 |
| **B. Post-treatment (2019–2023)** | | | |
| U.S. Federal Funding Amount | 7.820 | 8.159 | < 0.001 |
| U.S. Funding Amount | 7.917 | 8.255 | < 0.001 |
| Total Funding Amount | 8.174 | 8.396 | 0.005 |
| Pivot Size | 0.527 | 0.518 | 0.001 |
| Career Age | 26.900 | 26.799 | 0.476 |
| # Papers | 1.688 | 1.778 | < 0.001 |
| Average Normalized Citations | 0.453 | 0.453 | 0.933 |
| # New U.S. Collaborators | 1.868 | 2.076 | < 0.001 |
| Number of Observations | 6,926 | 13,685 | |

Table 2. Baseline Regression Results. The impact of U.S.-China geopolitical tensions on funding amount and pivot size. *** $p < 0.001$, ** $p < 0.05$, * $p < 0.1$. Standard errors, clustered at the individual level, are reported in parentheses.

| | (1) | (2) | (3) | (4) | (5) | (6) | (7) | (8) | (9) | (10) |
|---|---|---|---|---|---|---|---|---|---|---|
| | U.S. Federal Funding Amount | U.S. Federal Funding Amount | U.S. Funding Amount | Total Funding Amount | Pivot Size | Pivot Size | # U.S. Federal Grants | # U.S. Grants | # Total Grants | Reference Overlap |
| DID | -0.265*** | -0.193** | -0.188** | -0.114* | 0.007*** | 0.010*** | -0.017** | -0.016** | -0.013** | -0.015*** |
| | (0.061) | (0.060) | (0.060) | (0.060) | (0.001) | (0.001) | (0.006) | (0.006) | (0.006) | (0.001) |
| Career Age | | -0.069*** | -0.077*** | -0.078*** | | 0.006*** | -0.008*** | -0.009*** | -0.010*** | 0.005*** |
| | | (0.006) | (0.006) | (0.006) | | (0.000) | (0.001) | (0.001) | (0.001) | (0.000) |
| # Papers | | 0.827*** | 0.827*** | 0.745*** | | -0.014*** | 0.092*** | 0.093*** | 0.087*** | 0.035*** |
| | | (0.027) | (0.027) | (0.028) | | (0.001) | (0.003) | (0.003) | (0.003) | (0.001) |
| Average Cf | | -0.062** | -0.063** | -0.125*** | | -0.035*** | -0.009** | -0.009** | -0.015*** | 0.024*** |
| | | (0.029) | (0.030) | (0.031) | | (0.001) | (0.003) | (0.003) | (0.003) | (0.001) |
| # New U.S. Collaborators | | 0.007 | 0.014 | 0.034** | | 0.022*** | 0.002* | 0.003** | 0.005*** | -0.044*** |
| | | (0.013) | (0.013) | (0.014) | | (0.000) | (0.001) | (0.001) | (0.001) | (0.001) |
| Constant | 8.482*** | 8.287*** | 8.588*** | 8.863*** | 0.473*** | 0.346*** | 0.801*** | 0.842*** | 0.884*** | 0.190*** |
| | (0.026) | (0.146) | (0.145) | (0.147) | (0.001) | (0.005) | (0.015) | (0.015) | (0.015) | (0.005) |
| Year Fe | Y | Y | Y | Y | Y | Y | Y | Y | Y | Y |
| Individual Fe | Y | Y | Y | Y | Y | Y | Y | Y | Y | Y |
| N | 206110 | 206110 | 206110 | 206110 | 190619 | 190619 | 206110 | 206110 | 206110 | 190619 |
| Adj. R² | 0.697 | 0.702 | 0.698 | 0.683 | 0.634 | 0.642 | 0.771 | 0.771 | 0.762 | 0.452 |

**Table 3.** Moderating Effects. *** p < 0.001, ** p < 0.05, * p < 0.1. Standard errors, clustered at the individual level, are reported in parentheses.

|  | (1) | (2) | (3) | (4) | (5) | (6) |
|---|---|---|---|---|---|---|
|  | U.S. Federal Funding Amount | U.S. Funding Amount | Total Funding Amount | # U.S. Federal Grants | # U.S. Grants | # Grants |
| DID×Pivot Size | 0.685** | 0.665** | 0.610** | 0.089*** | 0.089*** | 0.083*** |
|  | (0.225) | (0.224) | (0.226) | (0.023) | (0.023) | (0.023) |
| DID | -0.525*** | -0.503*** | -0.403** | -0.063*** | -0.062*** | -0.055*** |
|  | (0.132) | (0.132) | (0.133) | (0.013) | (0.013) | (0.014) |
| Pivot Size | -0.272** | -0.281** | -0.294** | -0.013 | -0.015* | -0.017* |
|  | (0.095) | (0.095) | (0.095) | (0.009) | (0.009) | (0.009) |
| Career Age | -0.059*** | -0.067*** | -0.064*** | -0.007*** | -0.008*** | -0.009*** |
|  | (0.006) | (0.006) | (0.006) | (0.001) | (0.001) | (0.001) |
| # Papers | 0.778*** | 0.772*** | 0.738*** | 0.097*** | 0.098*** | 0.097*** |
|  | (0.030) | (0.029) | (0.030) | (0.003) | (0.003) | (0.003) |
| Average Cf | -0.054* | -0.060* | -0.067** | -0.000 | -0.001 | -0.001 |
|  | (0.032) | (0.032) | (0.033) | (0.003) | (0.003) | (0.003) |
| # New U.S. Collaborators | 0.021 | 0.028** | 0.030** | 0.003** | 0.004** | 0.004** |
|  | (0.013) | (0.014) | (0.014) | (0.001) | (0.001) | (0.001) |
| Constant | 8.527*** | 8.839*** | 8.920*** | 0.801*** | 0.844*** | 0.868*** |
|  | (0.161) | (0.160) | (0.161) | (0.017) | (0.017) | (0.017) |
| Year Fe | Yes | Yes | Yes | Yes | Yes | Yes |
| Individual Fe | Yes | Yes | Yes | Yes | Yes | Yes |
| N | 190619 | 190619 | 190619 | 190619 | 190619 | 190619 |
| Adj. $R^2$ | 0.696 | 0.692 | 0.683 | 0.767 | 0.767 | 0.763 |